# Antibiotic Resistance Microbiology Dataset (ARMD): A Resource for Antimicrobial Resistance from EHRs

Fateme Nateghi Haredasht[1†], Fatemeh Amrollahi[1], Manoj V. Maddali[2], Nicholas Marshall[3], Stephen P. Ma[4], Lauren N. Cooper[5], Andrew O. Johnson[6], Ziming Wei[7], Richard J. Medford[5,8], Sanjat Kanjilal[7], Niaz Banaei[9,10], Stanley Deresinski[9], Mary K. Goldstein[11], Steven M. Asch[12], Amy Chang[9], Jonathan H. Chen[1,4,13]

[1]Stanford Center for Biomedical Informatics Research, Stanford University, Stanford, CA, USA; [2]Division of Pulmonary and Critical Care Medicine, Department of Medicine, Stanford University School of Medicine, Stanford, CA, USA; [3]Division of Pediatric Infectious Diseases, Department of Pediatrics, Stanford University School of Medicine, Palo Alto, CA, USA; [4]Division of Hospital Medicine, Stanford University, CA, USA; [5]Clinical Informatics Center, University of Texas Southwestern Medical Center, Dallas, TX, USA; [6]Information Services, East Carolina University, Greenville NC, USA; [7]Department of Population Medicine, Harvard Medical School and Harvard Pilgrim Healthcare Institute, Boston, Massachusetts, USA; [8]Brody School of Medicine, Department of Internal Medicine, East Carolina University, Greenville NC, USA; [9]Division of Infectious Diseases and Geographic Medicine, Stanford University School of Medicine, Stanford, CA, USA; [10]Department of Pathology, School of Medicine, Stanford University, Palo Alto, CA, USA; [11]Department of Health Policy, Stanford University School of Medicine, Stanford, CA, USA; [12]Division of Primary Care and Population Health, Stanford University School of Medicine, Stanford, CA, USA; [13]Clinical Excellence Research Center, Stanford University, CA, USA

[†]Corresponding Author email: fnateghi@stanford.edu

## Abstract

The Antibiotic Resistance Microbiology Dataset (ARMD) is a de-identified resource derived from electronic health records (EHR) that facilitates research in antimicrobial resistance (AMR). ARMD encompasses big data from adult patients collected from over 15 years at two academic-affiliated hospitals, focusing on microbiological cultures, antibiotic susceptibilities, and associated clinical and demographic features. Key attributes include organism identification, susceptibility patterns for 55 antibiotics, implied susceptibility rules, and de-identified patient information. This dataset supports studies on antimicrobial stewardship, causal inference, and clinical decision-making. ARMD is designed to be reusable and interoperable, promoting collaboration and innovation in combating AMR. This paper describes the dataset's acquisition, structure, and utility while detailing its de-identification process.

## Background & Summary

Antimicrobial resistance (AMR) has emerged as a critical global health threat, compromising the effectiveness of antibiotics and leading to increased morbidity and mortality. In 2019, AMR was associated with nearly 5 million deaths worldwide, with at least 1.27 million directly attributable to resistant infections [1,2]. In the United States alone, over 2.8 million antimicrobial-resistant infections occur annually, resulting in more than 35,000 deaths[3]. AMR occurs when microorganisms such as bacteria, viruses, fungi, and parasites evolve mechanisms to withstand the effects of antimicrobial agents[4,5]. It also encompasses the selection and proliferation of organisms that inherently resist specific treatments, even without prior antimicrobial exposure. This problem is exacerbated by the overuse and misuse of antibiotics in clinical, agricultural, and community settings, which creates selective pressure favoring resistant strains.

Efforts to combat AMR require robust data resources to better understand resistance patterns, evaluate clinical practices, and develop evidence-based recommendations or practices for antimicrobial stewardship. However, comprehensive datasets that integrate microbiological and clinical data are rare. Moreover, the dynamic nature of resistance development necessitates datasets that capture temporal trends and patient-specific factors influencing AMR. Real-world data from electronic health records (EHR) offer a valuable opportunity to address this gap by providing granular information on microbial cultures, patient characteristics, and treatment outcomes[6–9]. Yet, creating meaningful and reliable datasets from EHR data presents several challenges, including heterogeneity in data representation, the need for rigorous de-identification, and ensuring data quality and interpretability.

Several publicly available datasets facilitate the study of AMR by providing genomic, phenotypic, and epidemiological insights. Resources such as the National Database of Antibiotic Resistant Organisms (NDARO)[10] and the Comprehensive Antibiotic Resistance Database (CARD)[11] focus on genetic determinants of resistance, while platforms like NARMS Now[12] and ResistanceMap[13] offer population-level surveillance data. Additionally, datasets like AMR-UTI[14] provide clinical insights from urinary tract infections, though they often lack comprehensive metadata linking microbiological findings to patient care and treatment outcomes. While these datasets contribute significantly to antimicrobial resistance research, they often remain siloed in their focus, emphasizing either genetic markers, population-level surveillance, or isolated clinical findings.

The Antimicrobial Resistance Microbiology Dataset (ARMD) offers a uniquely integrated dataset that combines microbiological culture data, antibiotic susceptibility results, patient demographics, clinical history, and treatment exposures from a large, real-world hospital setting. By bridging laboratory and clinical data, ARMD enables deeper epidemiological analyses, supports the development of predictive models for empirical treatment strategies, and provides a foundation for studying antimicrobial stewardship in real-world healthcare environments. This dataset serves as a novel resource that facilitates both broad surveillance and detailed patient-level analyses to inform future AMR research and clinical decision-making.

The ARMD dataset further addresses key challenges in antimicrobial resistance research by providing a robust, longitudinal dataset derived from de-identified EHR data at Stanford Health Care. Spanning multiple years and including over 280,000 unique patients, ARMD captures a diverse patient population and integrates microbiological, clinical, and longitudinal patient-level data to create a comprehensive resource for studying antimicrobial resistance patterns. The microbiological data within ARMD includes detailed information on microbiology specimens, such as body site, organism identification, and antibiotic susceptibility profiles. Unlike many existing datasets, ARMD also includes records of negative cultures, which serve as valuable indicators for assessing disease severity, estimating treatment success or failure, and understanding patterns of microbial clearance over time.

ARMD's comprehensive structure supports a wide range of research applications. It enables trend analysis, facilitating the monitoring of temporal shifts in resistance patterns across different organisms and clinical settings. The dataset is also valuable for risk factor identification, allowing researchers to assess how demographic and clinical characteristics contribute to the development of resistant infections. Additionally, ARMD serves as a foundation for predictive modeling efforts, aiding in the development of machine learning algorithms to predict resistance emergence and optimize empiric antibiotic therapy [15,16]. The insights derived from ARMD can further inform policy development, guiding antimicrobial stewardship programs and public health strategies aimed at mitigating the spread of resistance[15,17–20]. Making this dataset openly accessible encourages collaboration and innovation in AMR research, supporting global efforts to tackle this urgent public health threat.

# Methods

## Data Acquisition

The ARMD dataset was developed using de-identified EHR from Stanford Health Care, encompassing a broad range of microbiological, clinical, and demographic data collected from 1999 to 2024. The dataset integrates microbiology laboratory results, demographic data, clinical encounters, antibiotic exposures, and socioeconomic indicators to enable comprehensive analyses of AMR patterns.

Stanford Health Care uses the Epic EHR system to manage patient records. Data from Epic's operational database (Chronicles)—which is optimized for real-time transactional processing—are regularly extracted into Clarity, Epic's relational database designed for reporting and research purposes. At Stanford Health Care, the Clarity database is built on an Oracle-based system. For research and data analysis purposes, data from Clarity is integrated into the STAnford medicine Research data Repository (STARR)[21], which serves as a centralized data lake[22]. Access to STARR data was granted under Stanford IRB approval with review and oversight by the Privacy Office and Hospital Compliance Office to ensure regulatory compliance and patient privacy. Data extraction for ARMD was conducted using structured SQL queries executed on STARR's BigQuery interface. From STARR, relevant data for ARMD—such as microbiological cultures, laboratory test results, vital signs, medication exposures, and patient demographics—were extracted. The extraction process utilized Google BigQuery, a managed, cloud-based data warehouse that enables fast and scalable querying of large datasets. Researchers accessed BigQuery through a secure Virtual Private Network (VPN) using Cisco technology, ensuring data privacy and compliance with institutional security protocols. Organism identification was performed using Matrix-Assisted Laser Desorption Ionization Time-of-Flight (MALDI-TOF) mass spectrometry (Bruker Biotyper). Antibiotic susceptibility testing was conducted using the Vitek2 instrument (bioMérieux) for blood and urine cultures and the MicroScan WalkAway system (Beckman Coulter) for respiratory cultures. Minimum inhibitory concentrations were interpreted based on Clinical & Laboratory Standards Institute (CLSI) breakpoints.

## Inclusion/Exclusion Criteria

The ARMD dataset includes both inpatient and outpatient cultures from adult patients (aged 18 years or older) with urine, blood, and respiratory cultures. These three culture types were selected due to their clinical significance in antimicrobial resistance research, representing common sites of bacterial infections. Fungal, viral, and parasitic cultures were not explicitly included, as ARMD primarily focuses on bacterial resistance. To enhance data relevance and minimize redundancy, repeated cultures from the same patient within a two-week period were excluded. The dataset includes both positive and negative culture results, with positivity determined by the identification of specific organisms.

## Data Processing & Transformation

Organism and antibiotic names were standardized to resolve inconsistencies caused by varying nomenclature or formatting. When explicit susceptibility results were unavailable, intrinsic resistance was determined using Clinical and Laboratory Standards Institute (CLSI) standards, and implied susceptibility was inferred by linking susceptibility results between related antibiotics using predefined Stanford Microbiology Lab protocols, which are also based on CLSI standards[23–25]. These rules, documented in the related file, were systematically applied across all records. For example, susceptibility to an earlier-generation cephalosporin (e.g., cefazolin) implied susceptibility to a later-generation cephalosporin (e.g., ceftriaxone) based on established microbiological principles. These rules, documented in the related file, were systematically applied across all records. All data were de-identified following the Safe Harbor method in accordance with the National Institute of Standards and Technology (NIST) guidelines. Additionally, the clinical text was anonymized using the TiDE algorithm to ensure compliance with privacy regulations[26]. De-identification was performed in compliance with the Health Insurance Portability and Accountability Act (HIPAA) and Stanford Health Care's privacy regulations. Specifically, demographic data were anonymized by replacing exact ages with predefined age bins (e.g., 18–24, 25–34) and grouping all patients aged 89 or older into a single "90+" category. Sex was anonymized as binary values (0 and 1), with no further specification of sex labels. All date and time fields, including culture order dates, laboratory test dates, and medication administration times, underwent temporal jittering. This process involves applying random offsets to time-related data at the patient level, obscuring exact dates while preserving the relative temporal relationships essential for longitudinal analyses. No statistical imputation was applied, ensuring that users of the dataset could handle missing data according to their specific research methodologies.

## Data Structure & Schema

The data are structured to reflect the clinical timeline relevant to microbiological culture collection, including patient-level factors, clinical data, and culture-specific results. Figure 1 illustrates the data flow and relationships among the various data elements, highlighting how patient demographics, healthcare exposures, clinical data, and microbiological findings are linked.

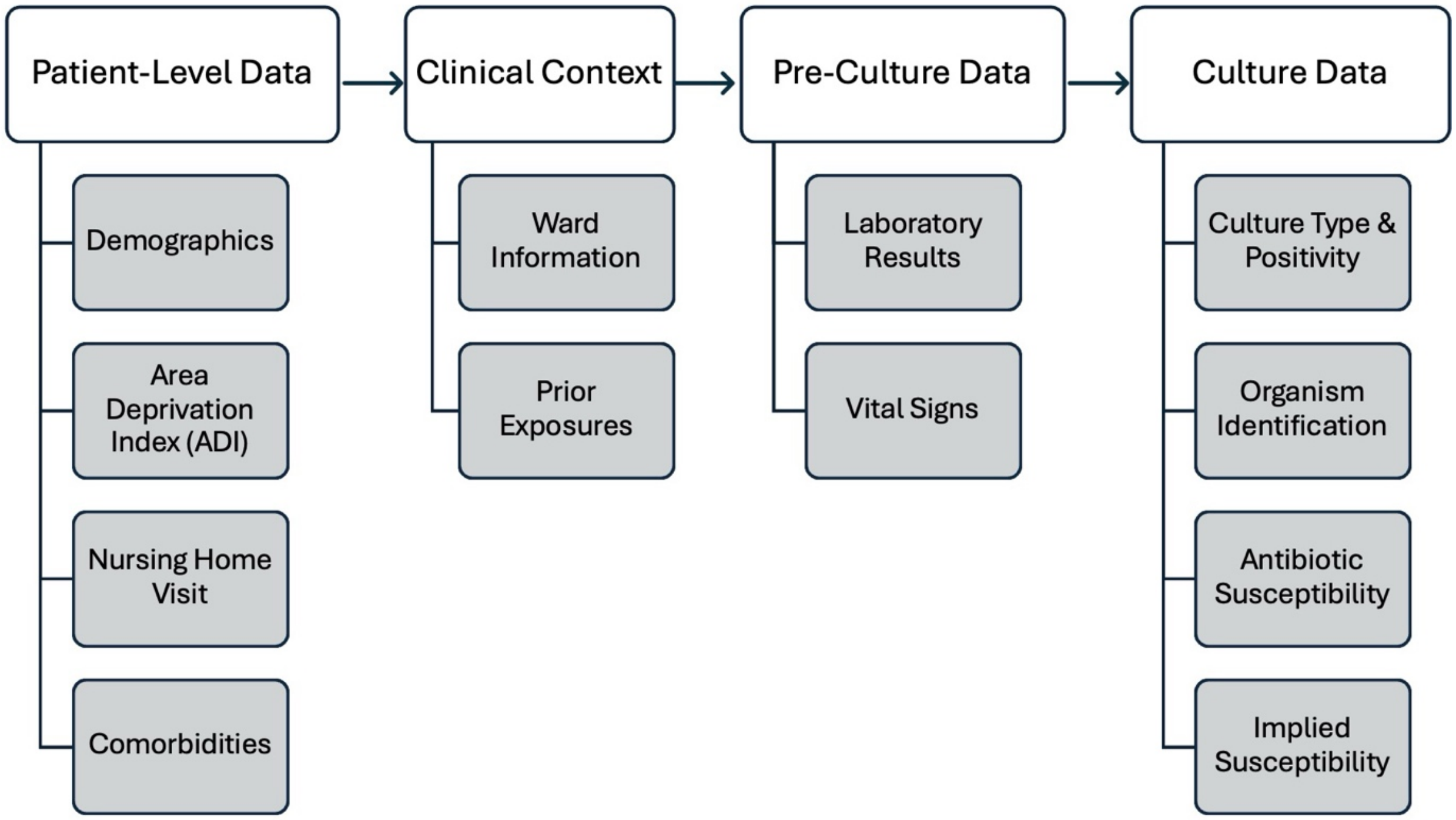

**Figure 1.** This flowchart highlights the connections between patient demographics, clinical data, microbiological cultures, and post-culture insights, enabling a longitudinal view of antimicrobial resistance.

The Patient-Level Data layer includes patient characteristics such as demographics (age, sex), socioeconomic indicators via the Area Deprivation Index (ADI), comorbidities (derived from the Elixhauser Comorbidity Index), and nursing home visits, which are known to confer antimicrobial resistance risk. The Clinical Context layer provides details about the care environment and relevant exposures surrounding culture collection, including ward information (e.g., intensive care unit [ICU], emergency department [ED], inpatient, outpatient) and prior exposures to antibiotics, medications, or procedures that may impact infection risk or resistance patterns. The Culture Collection layer focuses on the specific microbiological culture, capturing laboratory results (e.g., white blood cell counts, lactate levels) recorded within the 14 days preceding the culture order time and vital signs (e.g., heart rate, blood pressure, temperature) recorded within the 48 hours preceding the culture order. The Culture Data layer contains the results of the microbiological analysis, including culture type and positivity, organism identification, and antibiotic susceptibility profiles. It also includes implied susceptibility, inferred using established microbiological rules[23–25] to facilitate analysis when direct testing was not performed.

## Ethical Considerations

While the ARMD dataset has undergone rigorous de-identification processes, ethical data use remains paramount. Researchers should apply appropriate data security measures and respect the ethical guidelines outlined in the dataset's documentation. This study was approved by the Stanford University Institutional Review Board (IRB #70466). The IRB granted a waiver of patient consent in accordance with 45 CFR 164.512(i)(2)(ii).

## Data Records

### Variables and attributes

The ARMD dataset[27] is available at Dryad and encompasses a wide range of variables that are organized into multiple linked tables, each offering a unique perspective on a patient's microbiological, demographic, and clinical characteristics. To facilitate downstream analyses, the dataset includes tables on implied antibiotic susceptibility relationships and rules applied for inferring susceptibility where direct testing was not available. Researchers can also leverage longitudinal data, capturing the timing of infections, prior medical procedures, and medication exposures relative to culture orders, enabling temporal analyses.

At the core of ARMD is the microbiological cultures cohort, which includes details about culture types—urine, respiratory, and blood cultures—along with the identified organisms and their antibiotic susceptibilities. Antibiotic susceptibility results were included for 55 antibiotics and categorized into five groups: susceptible, resistant, intermediate, inconclusive, and synergism. Synergism refers to cases where the interaction between two antibiotics results in an enhanced effect, meaning the combined treatment is more effective than either antibiotic alone. This category captures instances labeled as "Synergy" or "No Synergy" in the dataset. Additional features include the culture's ordering mode (inpatient or outpatient) and the order's timing.

The dataset situates each culture event within its clinical context. The ward information provides insights into the care environment where cultures were collected, distinguishing between inpatient wards, intensive care units (ICU), emergency departments (ED), and outpatient clinics.

To capture potential influences on culture outcomes, ARMD includes records of prior antibiotic exposures. This component details the antibiotic name, class, and subtype, enabling analyses of how previous treatments may affect organism susceptibility and resistance development. The timing of these exposures relative to culture collection is recorded, supporting studies on the impact of prior antibiotic use on resistance development. Additionally, the dataset tracks microbial resistance trends on both individual and population levels over time, recording the evolution of resistance relative to culture events for specific organisms and antibiotics. Historical infection data are captured through the inclusion of a prior infecting organism table, which documents organisms identified in previous cultures for each patient. This enables longitudinal analyses of infection recurrence and its potential influence on current antimicrobial resistance. The table records the identified organism and the timing of the prior infection relative to each collected culture.

Patient demographics offer an essential context for stratifying analyses by age (binned into predefined ranges) and sex (binary-coded). In addition, the dataset incorporates socio-environmental factors through the inclusion of ADI scores, which capture neighborhood-level socioeconomic characteristics based on patient ZIP codes from the Neighborhood Atlas[28]. ADI scores designed for 9-digit ZIP codes account for factors such as income, education, employment, and housing quality, providing a broader context for understanding disparities in AMR risk. For records with only 5-digit ZIP codes, missing ADI scores were replaced with the average ADI score calculated from 9-digit ZIP codes sharing the same first 5 digits. For other cases with invalid or unavailable ADI scores (e.g., marked as P, U, or NA), no imputation was performed, and these entries were left as null values in the dataset.

Recognizing the role of long-term care facilities in AMR dynamics, nursing home visits are also documented, specifying the number of days between visits and culture orders, up to 90 days, to highlight potential risk factors for resistant infections.

Comprehensive laboratory data are integrated into the dataset, capturing key clinical measurements taken around the time of each culture order. Variables include white blood cell count, hemoglobin, creatinine, lactate, and procalcitonin, among other routinely collected studies. Each metric is summarized using statistical descriptors such as medians, quartiles (Q25, Q75), and first and last recorded values. Furthermore, vital sign data—including heart rate, blood pressure, temperature, and respiratory rate—provide additional clinical context, enabling analyses of physiological responses to infection.
Comorbid conditions are mapped using standardized indices such as the Elixhauser Comorbidity Index[29] and the Agency for Healthcare Research and Quality (AHRQ) Clinical Classifications Software Refined (CCSR)[30]. Each comorbidity is timestamped relative to the culture. Notably, ongoing comorbidities are flagged using NULL values in the end date field, indicating that the condition was active at the time of culture collection. These NULL values do not represent missing data or the absence of the condition. Additionally, procedural history is also provided, with records of medical procedures (e.g., central venous catheter placements, mechanical ventilation) performed prior to culture orders, derived from Current Procedural Terminology (CPT) codes.

Lastly, the implied susceptibility table infers antibiotic susceptibility for drugs not tested using an extensive set of predefined rules. This table captures cases where susceptibility to one antibiotic can imply susceptibility or resistance to another, based on established microbiological and pharmacological principles. The table is designed to enhance the interpretability of susceptibility data by incorporating implied relationships between antibiotics, which can be critical for guiding clinical decision-making and understanding resistance patterns. Additionally, we share the rules applied to derive these implied relationships, providing transparency and enabling researchers to understand and reproduce the logic behind the inferred data. This derived table leverages microbiological principles to capture relationships between antibiotics.

## Demographics and Microbiological Culture Data

ARMD comprises 751,075 microbiological culture records collected from 283,715 unique patients. Urine cultures constitute the majority of samples (50.0%), blood cultures represent 38.8%, and respiratory cultures account for 11.3%. The dataset spans from December 1999 to February 2024; however, there is a noticeable increase in recorded culture orders starting in 2008. This shift aligns with Stanford's adoption of Epic as the EHR system, which significantly improved data collection and documentation.

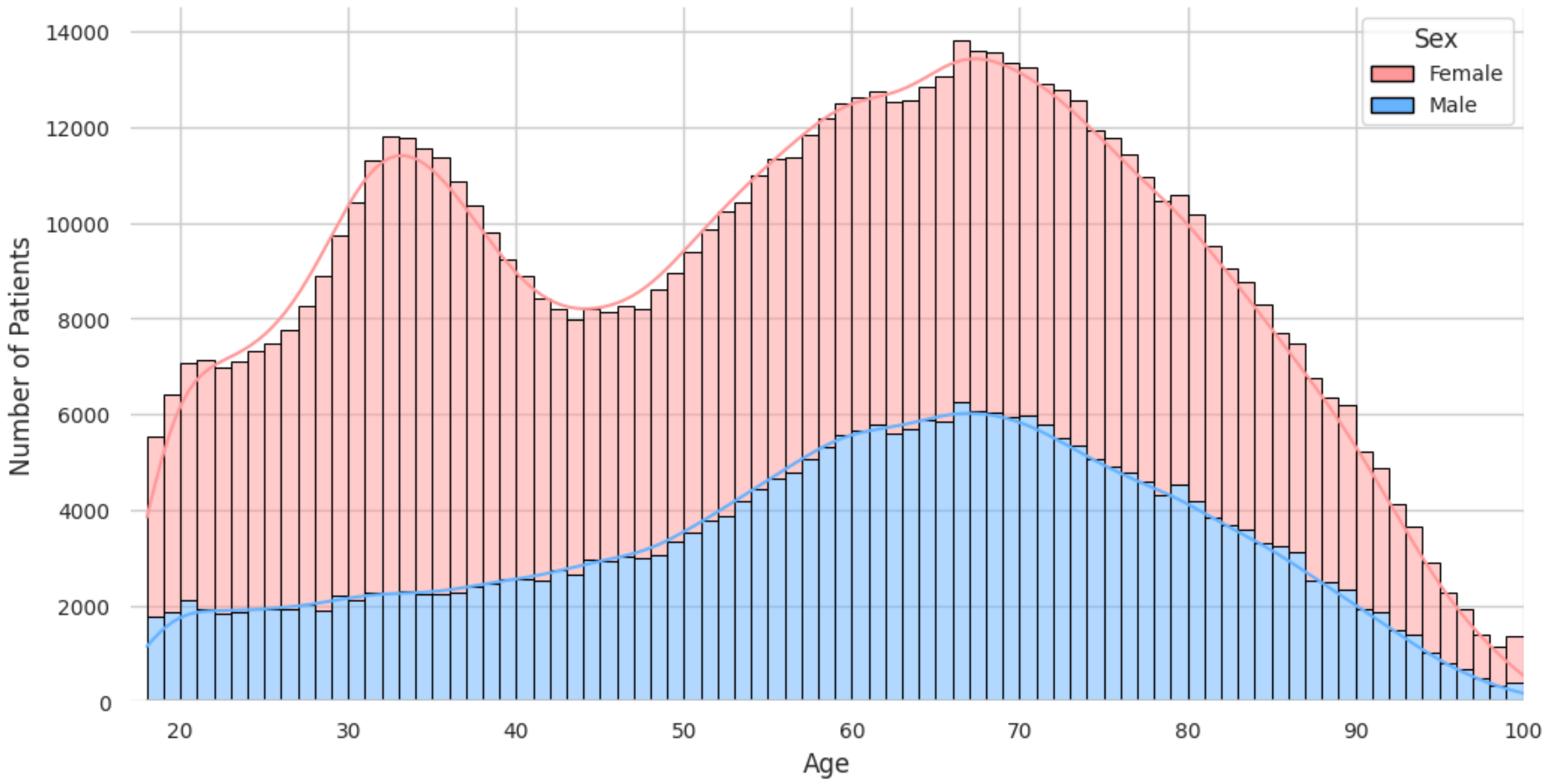

**Figure 2.** A histogram showing the age distribution of patients within the ARMD dataset.

The patient population demonstrates a broad age distribution, as illustrated in Figure 2, with an average age of 56.7 years. The sex distribution within the cohort reveals a predominance of female patients, accounting for 66.9% (189,864 patients) of the total population, and male patients form 33.0% (93,763 patients), while a minimal fraction (0.03%, n = 82) have an unknown sex designation.

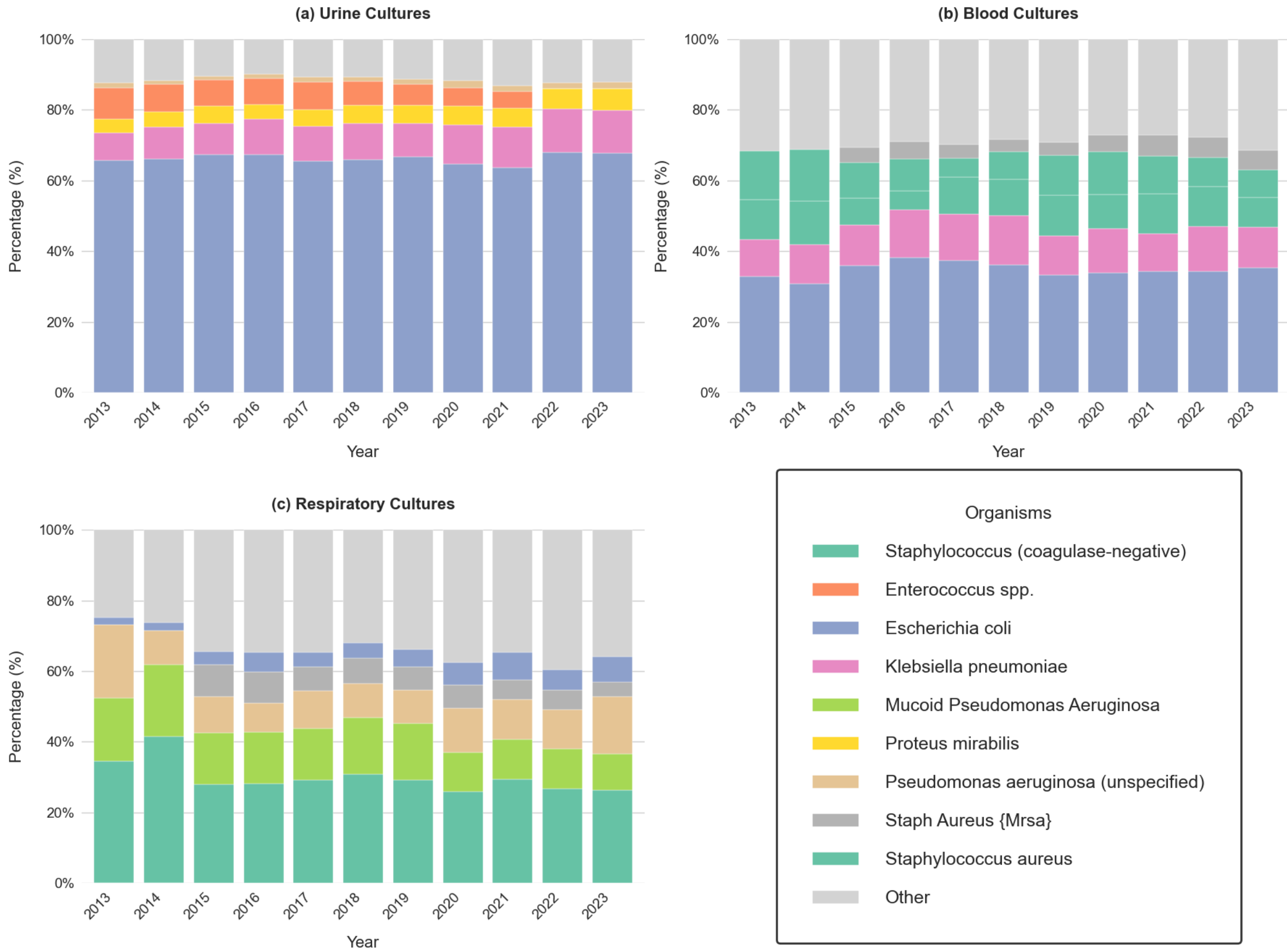

**Figure 3.** Distribution of the top five most common bacterial organisms identified in urine, blood, and respiratory cultures over time (2013–2023). The stacked bar charts show the relative percentage of each organism by year, with an additional "Other" category aggregating all less frequent isolates. The organisms show different prevalence patterns across culture types, reflecting variations in infection sources and microbial ecology.

Figure 3 presents the annual distribution of the top five most common organisms identified in urine, blood, and respiratory cultures from 2013 to 2023. In urine cultures (Figure 3-a), *Escherichia coli (E. coli)* is the predominant pathogen, consistently accounting for more than 60% of isolates. *Klebsiella pneumoniae* and *Proteus mirabilis* are the next most frequently detected organisms, with little variation over time. This stability in distribution indicates a consistent microbiological profile for UTIs within the cohort, consistent with established epidemiological trends nationwide [31–33].

In blood cultures (Figure 3-b), a more diverse range of pathogens is observed compared to urine cultures. While *E. coli* remains the most common pathogen, Staphylococcus aureus and coagulase-negative staphylococci are more prevalent, reflecting the tendency of gram positive cocci to cause bloodstream infections.

In respiratory cultures (Figure 3-c), *Pseudomonas aeruginosa* is the most frequently isolated pathogen, possibly related to selection bias among patients who underwent respiratory culture testing from either non-invasive (e.g., induced sputum) or invasive (e.g., bronchoalveolar lavage) methods. A distinction between mucoid and non-mucoid *Pseudomonas aeruginosa* is observed, likely reflecting changes in microbiology reporting standards. Mucoid strains are clinically significant, particularly in chronic respiratory infections. Other notable organisms include *Klebsiella pneumoniae* and *Staphylococcus aureus*, both of which remain stable contributors to respiratory infections throughout the study period.

## Technical Validation

To maintain data integrity, we minimized structural changes during transformation, preserving the clinical semantics of the original EHR data. Organism names, antibiotic identifiers, and susceptibility labels were standardized across records to resolve inconsistencies arising from varied reporting practices over the dataset's 15-year span. Although the dataset spans from 1999 to 2024, structured and consistent EHR documentation began with the adoption of Epic in 2008. Therefore, most validations focus on data from the 15-year period beginning in 2008.

To validate dataset completeness and accuracy, we performed internal cross-checks on key variables, including culture positivity, organism-antibiotic susceptibility pairings, and linkage across patient demographics, clinical history, and laboratory results. Descriptive statistics were computed to assess expected distributions of age, sex, culture types, and pathogen prevalence, all of which aligned with established epidemiological benchmarks.

Version-controlled scripts were used throughout the data processing pipeline. All transformation logic, including implied susceptibility rules, is documented and provided alongside the dataset to support reproducibility and community validation. Issues identified during development were tracked collaboratively and resolved through iterative testing with domain experts in infectious diseases, clinical microbiology, and clinical informatics.

## Usage Notes

This dataset supports studies in several critical areas, including AMR trend analysis, the development of predictive models for empirical antibiotic selection, and the examination of clinical and environmental factors that influence resistance patterns. The inclusion of granular data on culture positivity, organism identification, antibiotic susceptibility, prior medication exposures, comorbidities, and nursing home visits allows for detailed epidemiological analyses and modeling of resistance risk factors. Resistant isolates were not biobanked by the Stanford Health Care clinical microbiology laboratory and are not available for external laboratory access. A README file is included in the Dryad repository to guide users in navigating the dataset structure and contents. While regular updates are not planned, future revisions will be versioned and documented in the repository.

### Handling Missing Data

Empty fields within the dataset are explicitly marked as "null" to maintain clarity. Users are advised to handle these values appropriately during analysis, particularly when conducting statistical modeling or machine learning tasks.

## Code Availability

No pre-packaged analysis scripts are included with the dataset. However, the structured format of the CSV files supports seamless integration with common data analysis tools, such as Python (pandas, scikit-learn, etc), R, SPSS, or SAS. Users requiring guidance on analytical workflows can contact the dataset authors for further support.

## Acknowledgments

This work was supported by the National Institute of Allergy and Infectious Diseases (NIAID) under NIH R01 Grant R01AI179155 from the National Institutes of Health. We also acknowledge the contributions of our collaborators at Stanford University, UT Southwestern Medical Center, Harvard Medical School, and Harvard Pilgrim Healthcare Institute for their support in data harmonization, validation, and scientific guidance throughout this project.

## Author Contributions

F.NH., F.A., and M.V.M. built the ARMD dataset. All authors gave input into the database development process and contributed to writing the paper.

## Competing Interests

J.H.C. reported being a co-founder of Reaction Explorer LLC, develops and licenses organic chemistry education software, paid consulting fees from Sutton Pierce, Younker Hyde MacFarlane, and Sykes McAllister as a medical expert witness, and paid consulting fees from ISHI Health. F.NH. reported being paid consulting fees from ISHI Health. The remaining authors declare no competing interests.